\documentstyle[11pt,epsfig]{article}
\def\itemitem{\par\indent\hangindent=2\parindent\textindent}                    
\def\dis{\displaystyle}

\begin{document}

\begin{center}
\LARGE\bf Towards the Light Front Variables\\
for High Energy Production Processes.\\
\end{center}
\vspace{1cm}

\begin{center}
\bf N.S. Amaglobeli$^1$, S.M. Esakia$^1$, \underbar{\bf V.R. 
Garsevanishvili$^{2,3}$},\\
G.O. Kuratashvili$^1$, $\dagger$N.K. Kutsidi$^{1,4}$, R.A. Kvatadze$^{1,4}$,\\
Yu V. Tevzadze$^1$, T.P. Topuria$^4$\\
\end{center}

\itemitem{$^1$} High Energy Physics Institute, Tbilisi,
    State University, University Str. 9,
    380086 {Tbilisi, Rep. of Georgia}
\itemitem{$^2$} Laboratoire de Physique Corpusculaire,
    Universit\'e Blaise Pascal,
    63177 {Aubi\`ere Cedex, France}
\itemitem{$^3$} Mathematical Institute of the Georgian
    Academy of Sciences, M. Aleqsidze Str. 1, 
    380093 {Tbilisi, Rep. of Georgia}
\itemitem{$^4$} Joint Institute for Nuclear Research,
    {141980 Dubna, Russia}

\begin{center}
\begin{tabular}{l}
Contact author : {V.R. Garsevanishvili}\\
E-mail: garse@clrvax.in2p3.fr (till May 5, 1997)\\
\hphantom{E-mail: }garse@imath.acnet.ge (after May 5, 1997)\\
Fax: +33.473.264598 (till May 5, 1997)\\
\hphantom{Fax: }+ 995.32.990689 (after May 5, 1997)\\
\end{tabular}
\end{center}
\vspace{1cm}

\begin{center}
\large\bf ABSTRACT\\
\end{center}

Scale invariant presentation of inclusive spectra in terms of light front
variables is proposed. The variables introduced go over to the well-known 
scaling
variables $x^{}_F=2p^{}_z/\sqrt{s}$ and $x^{}_T=2p^{}_T/\sqrt{s}$ in the high
$p^{}_z$ and high $p^{}_T$ limits respectively.

Some surface is found in the phase space of produced 
$\pi^{\pm}$-mesons in the inclusive reaction $\bar pp\to\pi^{\pm}X$ at
22.4~GeV/c, which separates two groups of particles with significantly
different characteristics.
In one of these regions a naive statistical model seems to be
in a good agreement with data, whereas it fails in the second region.
\bigskip

\noindent{\bf Classification codes:} PACS numbers: 13.85 Ni, 13.90+i

\noindent{\bf Key words}: Light front, inclusive, hadron-hadron, 
electron-positron, relativistic heavy ions, deep inelastic.
\vfill

\clearpage

The study of single particle inclusive processes \cite{1} remains one of the
simplest and effective tools for the investigation of multiple production of
secondaries at high energies. The consequences of the
limiting fragmentation hypothesis \cite{2} and those of the parton model
\cite{3} and the principle of automodelity for strong interactions \cite{4}
have been formulated in this way.

An important role in establishing of many properties of multiple production
is played by the choice of kinematic variables in terms of which observable
quantities are presented (see in this connection, e.g. \cite{5,6,7}). The
variables which are commonly used are the following:\hfil\break
 the Feynman 
$x^{}_F=2p^{}_z/\sqrt{s}$,
 rapidity $y=\dis {1\over{2}}{\rm ln}
[(E+p^{}_z)/(E-p^{}_z)]$,
transverse scaling variable $x^{}_T=2p^{}_T/\sqrt{s}$ etc.
In the case of azimuthal symmetry the surfaces of const $x^{}_F$ are the 
planes $p^{}_z=x^{}_F\sqrt{s}/2$, surfaces of constant $y$ are the
hyperboloids
\begin{eqnarray*}
p^2_z\left[\left({1+e^{2y}\over{1-e^{2y}}}\right)^2-1\right]-p^2_T=m^2
\end{eqnarray*}

\noindent and the surfaces of constant $x^{}_T$ are the straight lines 
$p^{}_T=x^{}_T\sqrt{s}/2$ in the phase space.

Here we propose a unified scale invariant variable for the presentation of single
particle inclusive distributions, the properties of which are described
below.

Consider an arbitrary 4--momentum $p^{}_{\mu}(p^{}_0,\vec p)$ and introduce the light
front combinations \cite{8}:
\begin{eqnarray}
p^{}_{\pm}=p^{}_0\pm p^{}_3 \label{eq1}
\end{eqnarray}

If the 4--momentum $p^{}_{\mu}$ is on the mass shell $(p^2=m^2)$, the combinations
$p^{}_{\pm},\ \vec p^{}_T$ (where $\vec p^{}_T=(p^{}_1,\ p^{}_2)$) define the so called
horospherical coordinate system (see, e.g. \cite{9,10}) on the 
corresponding mass shell hyperboloid
$p^2_0-\vec p\, ^2=m^2$. Corresponding hyperboloid in the velocity space is
the realization of the curved space with constant negative curvature, i.e.
the Lobachevsky space.

Let us construct the scale invariant variables:
\begin{eqnarray}
\xi^{\pm}=\pm {p^c_{\pm}\over{p^a_{\pm}+p^b_{\pm}}} \label{eq2}
\end{eqnarray}

\noindent in terms of the 4--momenta $p^a_{\mu},\ p^b_{\mu},\ p^c_{\mu}$ of particles
$a,\ b,\ c$, entering the inclusive reaction $a+b\to c+X$. The
$z$-axis is taken to be the collision axis, i.e. $p^{}_z=p^{}_3=p^{}_L$. Particles $a$
and $b$ can be hadrons, heavy ions, leptons. Note that the use of similar 
variables
turned out to be successful in theoretical studies of relativistic composite systems
(see, e.g. \cite{11}--\cite{25}), in the theoretical and experimental
studies of nuclear reactions with beams of relativistic nuclei (see, e.g.
[22,26,27]) and in the study of quark confinement in QCD (see, e.g. \cite{28}).
Combinations like Eq.(\ref{eq1}) appear also when
considering the scale transformations \cite{29} in the theory with
fundamental length (see, e.g. \cite{30}).

Invariant differential cross section in terms of $(\xi^{\pm},\ \vec p^{\, c}_T)$
- variables looks as follows (assuming the azimuthal symmetry):
\begin{eqnarray}
E^c{d\sigma\over{d\vec p^c}}={|\xi^{\pm}|\over{\pi}}\ {d\sigma\over{
d\xi^{\pm}dp^{c^2}_T}} \label{eq3}
\end{eqnarray}

It is interesting to note the properties of $\xi^{\pm}$ - variables in some
limiting cases. Let us choose the centre of mass frame, where:
\begin{eqnarray}
\xi^{\pm}&=&\pm {E^c\pm p^c_z\over{\sqrt{s}}}=\pm {E^c+|p^c_z|\over{
\sqrt{s}}}\ ; \label{eq4}\\
E^c&=&\sqrt{p^{c^2}_z+p^{c^2}_T+m^{c^2}} \nonumber
\end{eqnarray}

The upper sign in Eq.(\ref{eq4}) is used for the right hand side hemisphere and
the lower sign for the left hand side hemisphere in the centre of mass
frame.

Consider two limiting cases:

\itemitem{1)} $|p^c_z|\gg p^c_T$ - fragmentation region, according to
the common terminology.

\itemitem{} In this case:
\begin{eqnarray}
\xi^{\pm}\longrightarrow {2p^c_z\over{\sqrt{s}}}=x^{}_F \label{eq5}
\end{eqnarray}

\itemitem{2)} $p^c_T\gg |p^c_z|$ - high $p^{}_T$-region.

\itemitem{} In this case:
\begin{eqnarray}
\xi^{\pm}\longrightarrow {m^c_T\over{\sqrt{s}}}\longrightarrow
{p^c_T\over{\sqrt{s}}}={x^{}_T\over{2}}\ ;\ m^c_T=\sqrt{p_T^{c^2}+m^{c^2}}
\label{eq6}
\end{eqnarray}

Thus, in these two limiting regions of phase space $\xi^{\pm}$--variables
go over to the well known variables $x^{}_F$ and $x^{}_T$, which are
intensively used in high energy physics. 
$\xi^{\pm}$--variables are related to $x^{}_F$,\ $x^{}_T$ and $y$ as follows:
\begin{eqnarray}
\xi^{\pm}&=&{1\over{2}}\left(x^{}_F\pm \sqrt{x^2_F+x^2_{\perp}}\right)\ 
;\ x^{}_{\perp}={2m^c_T\over{\sqrt{s}}} \label{eq7}\\
y&=&\pm {1\over{2}}\ {\rm ln}\, {(\xi^{\pm}\sqrt{s})^2\over{m^{c^2}_T}}
\label{eq8}
\end{eqnarray}

The region $\displaystyle |\xi^{\pm}|<m^c/\sqrt{s}$ is
kinematically forbidden for the $\xi^{\pm}$--spectra integrated over all values
of $p^{c^2}_T$, and the region $\displaystyle |\xi^{\pm}|<
m^c_T/\sqrt{s}$ is forbidden for the $\xi^{\pm}$--spectra at fixed
values of $p^{c^2}_T$.

In the present paper we study the inclusive reaction $\bar pp \to\pi^{\pm} X$
at 22.4~GeV/c of the incident momentum. The details of the experiment can be
found in \cite{31}. In this case it is sufficient to study the right hand
side hemisphere only, due to the CP--symmetry of the reaction.

In Fig.~1a the $\xi^+$--distribution of $\pi^{\pm}$--mesons is shown.

$\xi^+$--distribution has two features, which makes it differ from the
corresponding $x^{}_F$--distribution:

\itemitem{1)} existence of the forbidden region near the point 
$\xi^{+}=0$ (cross section vanishes in the region 
$\displaystyle |\xi^{\pm}|<m^{}_{\pi}/\sqrt{s}$, 
\itemitem{2)} existence of maximum at some $\tilde{\xi}^+$ in the region of
relatively small $\xi^{+}$.

It is convenient to introduce the variable
\begin{eqnarray}
\zeta^+=-{\rm ln}\xi^+ \label{eq9}
\end{eqnarray}

\noindent in order to enlarge the scale in the region of small $\xi^+$. The maximum
at $\tilde{\zeta}^+$ is also observed in the invariant differential cross section
$\displaystyle {1\over{\pi}}\ {d\sigma\over{d\zeta^+}}$. However, the region
$\xi^+>\tilde{\xi}^+$ goes over to the region $\zeta^+<\tilde{\zeta}^+$ and vice
versa (see Fig.~1b).

In order to study the nature of this maximum we have investigated the
angular and $p^2_T$--distributions of $\pi^{\pm}$--mesons in the regions
$\xi^+<\tilde{\xi}^+(\zeta^+>\tilde{\zeta}^+)$ and
$\xi^+>\tilde{\xi}^+(\zeta^+<\tilde{\zeta}^+)$ separately. The results are presented
in Figs.~2a and 2b. The angular distribution of particles with 
$\xi^+>\tilde{\xi}^+(\zeta^+<\tilde{\zeta}^+)$ is sharply anisotropic in contrast
to the almost flat distribution of particles with
$\xi^+<\tilde{\xi}^+(\zeta^+>\tilde{\zeta}^+)$.
The slopes of $p^2_T$--distributions differ substantially.

Note, that the surfaces of constant $\xi^+$ are the paraboloids
\begin{eqnarray}
p^c_z={p^{c^2}_T+m^{c^2}-(\xi^+\sqrt{s})^2\over{-2\xi^+\sqrt{s}}}
\label{eq10}
\end{eqnarray}
\noindent in the phase space.
Thus the paraboloid
\begin{eqnarray}
p^c_z={p^{c^2}_T+m^{c^2}-(\tilde{\xi}^+\sqrt{s})^2\over{-2\tilde{\xi}^+
\sqrt{s}}} \label{eq11}
\end{eqnarray}

\noindent separates two groups of
particles with significantly different characteristics. 

It seems to be interesting to use $\xi^{\pm}$ and $\zeta^{\pm}$ variables in
deep inelastic 
electro - and weak production processes, in $e^+e^-$ - annihilation and in
relativistic heavy ion collisions (see in this connection recent
reviews
[32--38] and references therein) and to perform also event by event
analysis.

To describe the spectra in the region $\xi^+<\tilde{\xi}^+(\zeta^+>\tilde{\zeta}^+)$
the simplest statistical model (see, e.g. \cite{39}) with the Boltzman
$f(E)\sim e^{-E/T}$ and the Bose-Einstein $f(E)\sim (e^{E/T}-1)^{-1}$
distributions has been used.

The distributions $\displaystyle {1\over{\pi}}\ {d\sigma\over{d\zeta^+}}\ ,\ 
{d\sigma\over{dp^2_T}}$ and $\displaystyle {d\sigma\over{d\cos\theta }}$
look in this region as follows~:
\begin{eqnarray}
{1\over{\pi}}\ {d\sigma\over{d\zeta^+}}&\sim&\int_0^{p^2_{T,max}}Ef(E)dp^2_T,
\label{eq12a}\\
{d\sigma\over{dp^2_T}}&\sim&\int_0^{p^{}_{z,max}}f(E)dp^{}_z,\label{eq12b}\\
{d\sigma\over{d\cos\theta}}&\sim&\int_0^{p^{}_{max}}f(E)p^2dp,\label{eq12c}\\
E&=&\sqrt{\vec p\, ^2+m^2_{\pi}}\ ,\ \vec p\, ^2=p^2_z+p^2_T \label{eq12d}
\end{eqnarray}

\noindent where:
\begin{eqnarray}
p^2_{T,max}&=&(\xi^+\sqrt{s})^2-m^2_{\pi} \label{eq13a}\\
p^{}_{z,max}&=&{p^2_T+m^2_{\pi}-(\tilde{\xi}^+\sqrt{s})^2\over{
-2\tilde{\xi}^+\sqrt{s}}} \label{eq13b}\\
p^{}_{max}&=&{-\tilde{\xi}^+\sqrt{s}\cos\theta+\sqrt{(\tilde{\xi}^+\sqrt{s})^2-
m^2_{\pi}\sin^2\theta}\over{\sin^2\theta}} \label{eq13c}
\end{eqnarray}

The experimental distributions $\displaystyle {1\over{\pi}}\ 
{d\sigma\over{d\zeta^+}}\ ,\ {d\sigma\over{dp^2_T}}$ and $\displaystyle
 {d\sigma\over{d\cos\theta }}$ in the region $\xi^+<\tilde{\xi}^+(\zeta^+>
\tilde{\zeta}^+)$
have been fitted by Eqs. (\ref{eq12a}), (\ref{eq12b}) and (\ref{eq12c}),
respectively. The results of the fit given in Table~1 and Figs. 1b, 2a, 2b
show satisfactory agreement with experiment.

In the region $\xi^+>\tilde{\xi}^+(\zeta^+<\tilde{\zeta}^+)$ \  
$\zeta^+$--distribution
has been fitted by the formula:
\begin{eqnarray}
{1\over{\pi}}\ {d\sigma\over{d\zeta^+}}\sim (1-\xi^+)^n=(1-e^{-|\zeta^+|})^n 
\label{eq15}
\end{eqnarray}

\noindent and the $p^2_T$--distribution by the formula:
\begin{eqnarray}
{d\sigma\over{dp^2_T}}\sim \alpha e^{-\beta_1p^2_T}+(1-\alpha )
e^{-\beta_2p^2_T} \label{eq14}
\end{eqnarray}

Note that in the region $\xi^+\to 1$ the parameterization (19) goes over to the
well-known quark-parton model parameterization $(1-x )^n$ with 
$x =x^{}_F=2p^{}_z/\sqrt{s}$. The results of the fit are given in Table 2
and Figs. 1b and 2b. Since the dependence $(1-x )^n$ which is derived
for $x\to 1$ describes the data even in the region $x\to 0$ (where, in
general,
it must not work), but the dependence (\ref{eq15}) deviates from the data
in the region of small $\xi^+$, it seems
that the analysis of data in terms of $\xi^+$ and $\zeta^+$-distributions
is more sensitive to the multi-component models of multi-body production at
high energies.

Thus the spectra of $\pi^{\pm}$--mesons in the region $\xi^+<\tilde{\xi}^+
(\zeta^+>\tilde \zeta^+)$ are satisfactorily described by the formulae which follow
from the statistical model. The same formulae when extrapolated to the
region $\xi^+>\tilde{\xi}^+ (\zeta^+<\tilde \zeta^+)$ deviate
from the data. On the other hand, the dependence $(1-\xi^+)^n$ is in a good
agreement with data in the region $\xi^+>\xi^+(\zeta^+<\tilde{\zeta}^+)$
and deviates from them in the region $\xi^+<\tilde{\xi}^+(\zeta^+>
\tilde{\zeta}^+)$ (see Fig. 1b).

It is interesting to recall the similar situation in the study of black
body radiation, where the Wien formula describes the low frequency part
of the spectrum and does not describe the high frequency part, whereas
the situation is reversed in the case of Rayleigh-Jeans formula (see,
e,g, \cite{40}). To illustrate this in Fig. 3 the black body radiation 
intensity according to the Wien, Rayleigh-Jeans
and Planck formulae are plotted against the dimensionless variable $x =
\hbar\omega /kT$.

In conclusion, we feel that the use of the variables $\xi^{\pm}$ and
$\zeta^{\pm}$ can help to distinguish in between different dynamical
contributions, or test basic principles in other types of analysis, such
as two-particle correlations, HBT -- interferometry [41,42,43] and transverse
flow studies \cite{44}.

The authors are indebted to the staff of the two metre hydrogen bubble chamber
of JINR (Dubna) for supplying the experimental data. They would like to thank
{M. Jacob, L. Montanet, G. Roche, A.N. Tavkhelidze} for their kind interest
in this work and valuable discussions, {Z. Ajaltouni, J-P. Alard, 
A. Baldit, J. Bartke, A. Capella, I. Derado, P. Dupieux, B. Erazmus, 
H. Fonvieille,
P. Force, W. Geist, T. Hebbeker, R. Hong--Tuan, P. Juillot, M. Klein, C. Kuhn,
B. Levtchenko, J.-F. Mathiot,
C. Pienne, P. Pras,
P. Saturnini, T. Siemiarczuk, R. Stock, E. Strokovsky, M. Turala, J. Vary,
M. Winter} for useful 
discussions.
One of the authors ({V.R.G.}) expresses his deep gratitude 
to Guy {Roche}
and Bernard {Michel} for the warm hospitality at the Laboratoire de
Physique Corpusculaire, Universit\'e Blaise Pascal, Clermont-Ferrand,
to {V.G. Kadyshevsky, T.I. Kopaleishvili, H. Leutwyler, W. R\"uhl} for
supporting his stay at the L.P.C. and to
NATO for supporting this work.
\vfill
\clearpage

\begin{table}[htbp]
\caption{Results of the fits of $\displaystyle {1\over{\pi}}\ 
{d\sigma\over{d\zeta^+}}\ ,\ \displaystyle 
{d\sigma\over{d\cos\theta }}$
and $\displaystyle {d\sigma\over{dp^2_T}}$ --distributions in the
region $\xi^+<\tilde{\xi}^+(\zeta^+>\tilde{\zeta}^+)$.
\label{tab1}}
\begin{center}
\begin{tabular}{|c|c|c|c|c|}
\hline
 &\multicolumn{2}{c|}{}&\multicolumn{2}{c|}{}\\
 &\multicolumn{2}{c|}{T, GeV}&\multicolumn{2}{c|}{$\chi^2/N_{D.F.}$}\\
 &\multicolumn{2}{c|}{}&\multicolumn{2}{c|}{}\\
\hline
 &             &        &             & \\
 &Bose-Einstein&Boltzman&Bose-Einstein&Boltzman\\
 &             &        &             & \\
\hline
 &             &        &             & \\
$\displaystyle {1\over{\pi}}\ {d\sigma\over{d\zeta^+}}$&$0.134\pm 0.004$&$
0.119\pm 0.003$&10/8&12/8\\
 &             &        &             & \\
\hline
 &             &        &             & \\
$\displaystyle {d\sigma\over{d\cos\theta }}$&$0.091\pm 0.003$&$0.086\pm 
0.003$&16/7&15/7\\
 &             &        &             & \\
\hline
 &             &        &             & \\
$\displaystyle {d\sigma\over{dp^2_T}}$&$0.110\pm 0.001$&$0.105\pm 
0.001$&10/8&8/8\\
 &             &        &             & \\
\hline
\end{tabular}
\end{center}
\end{table}

\vfil

\begin{table}[htbp]
\caption{Results of the fits of $\displaystyle {d\sigma\over{d\zeta^+}}$
and $\displaystyle {1\over{\pi}}\ 
{d\sigma\over{dp^2_T}}$ --distributions in the
region $\xi^+>\tilde{\xi}^+(\zeta^+<\tilde{\zeta}^+)$.
\label{tab2}}
\begin{center}
\begin{tabular}{|c|c|c|c|c|c|}
\hline
 &        &         &         &   & \\
 &$\alpha$&$\beta_1$&$\beta_2$&$n$&$\chi^2 /N_{D.F.}$\\
 &        &(GeV/c)$^{-2}$&(GeV/c)$^{-2}$& & \\
 &        &         &         &   & \\
\hline
 &        &         &         &   & \\
$\displaystyle{1\over{\pi}}\ {d\sigma\over{d\zeta^+}}$& $-$ & $-$ & $-$ &$3.7\pm 
0.1$&7/9\\
 &        &         &         &   & \\
\hline
 &        &         &         &   & \\
$\displaystyle{d\sigma\over{dp^2_T}}$&$0.8\pm 0.03$&$6.0\pm 0.1$&$2.8\pm 
0.3$&\ &45/29\\
 &        &         &         &   & \\
\hline
\end{tabular}
\end{center}
\end{table}

\clearpage
\begin{twocolumn}
\begin{figure}[htbp]
\begin{center}\mbox{\epsfig{file=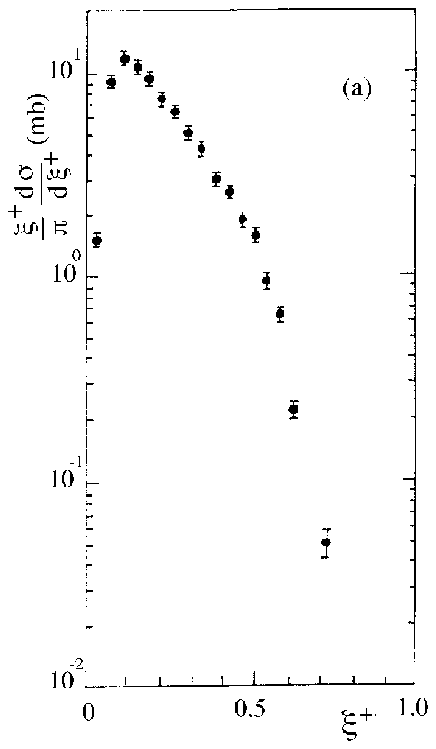,height=6cm}}Fig. 1\end{center}
\end{figure}
\nopagebreak
\begin{figure}[htbp]
\begin{center}\mbox{\epsfig{file=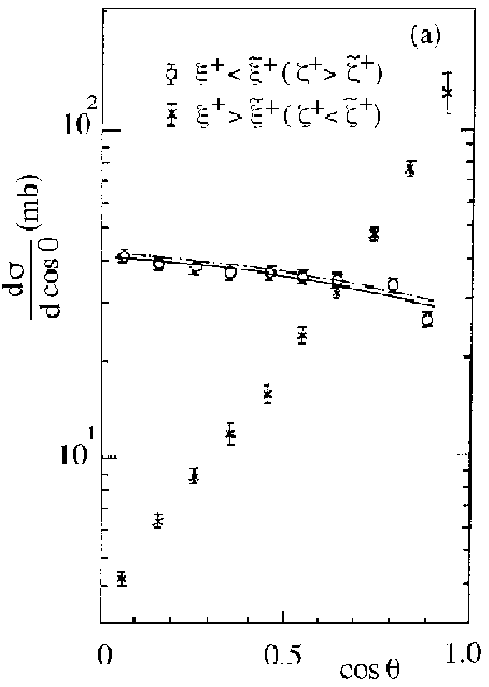,height=6cm}}Fig. 2\end{center}
\end{figure}
\nopagebreak
\begin{figure}[htbp]
\begin{center}\mbox{\epsfig{file=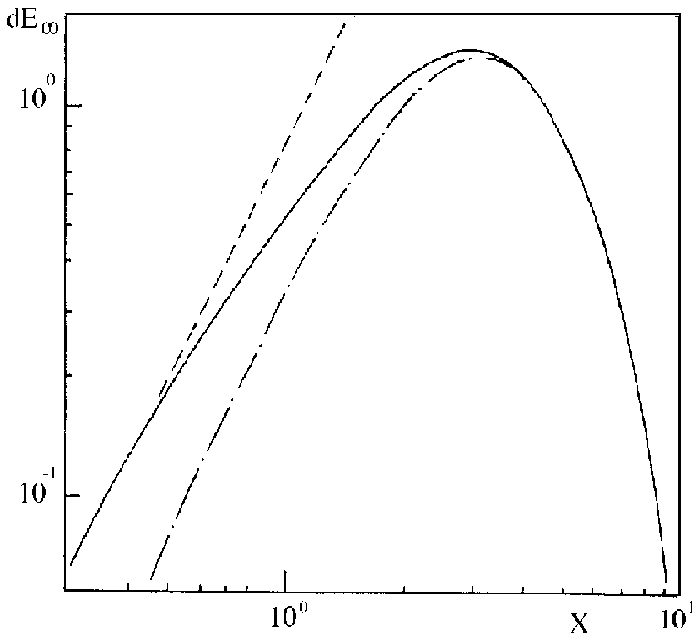,height=6cm}}Fig.~3\end{center}
\end{figure}

\newpage

\begin{figure}[htbp]
\begin{center}\mbox{\epsfig{file=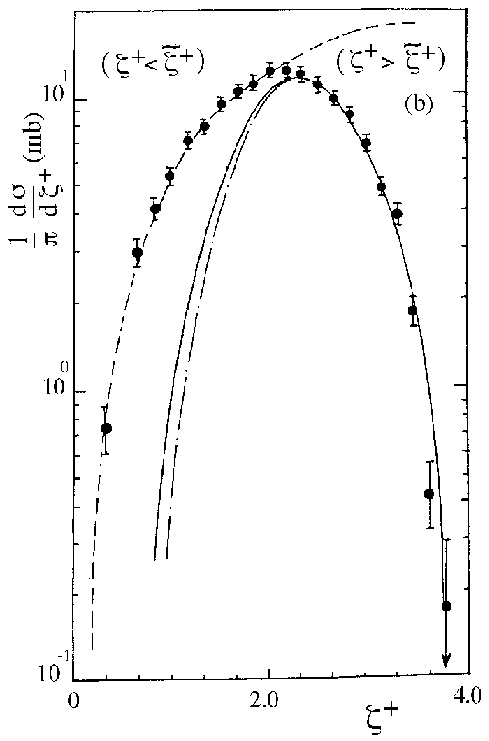,height=6cm}}\end{center}
\end{figure}
\nopagebreak
\begin{figure}[htbp]
\begin{center}\mbox{\epsfig{file=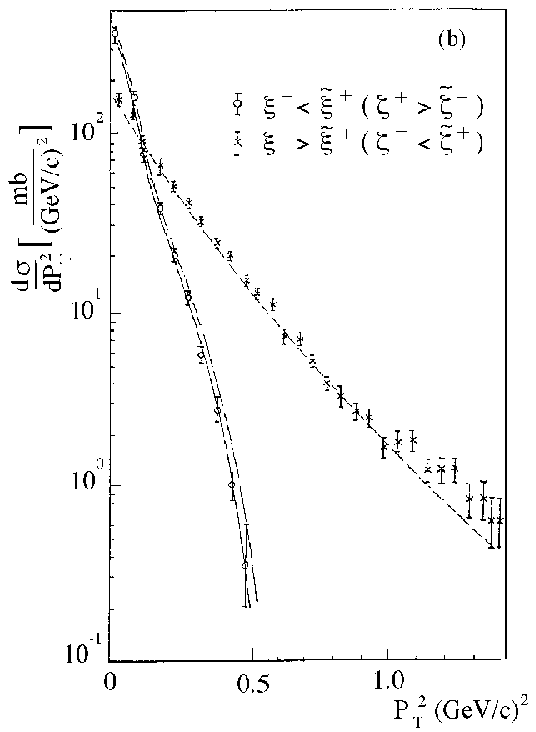,height=6cm}}\end{center}
\end{figure}

\end{twocolumn}

\clearpage
\onecolumn

\null

\noindent{\Large\bf Figure Captions}
\bigskip

\noindent Fig. 1: $\displaystyle {\xi^+\over{\pi}}\ {d\sigma\over{d\xi^+}}$ --
distribution of $\pi^{\pm}$ mesons in the reaction $\bar pp\to \pi^{\pm}X$
at\break  
22.4~GeV/c (1a). $\displaystyle {1\over{\pi}}\ {d\sigma\over{d\zeta^+}}$ 
-- distribution of $\pi^{\pm}$ mesons in the reaction 
$\bar pp\to\pi^{\pm}X$ at 22.4~GeV/c (1b),
\rule{1.3cm}{0.5mm} fit of the data in the region $\xi^+<\tilde{\xi}^+(\zeta^+>
\tilde{\zeta}^+)$ by the Bose-Einstein distribution, 
\rule{0.6cm}{0.5mm}$\,$.$\,$\rule{0.6cm}{0.5mm}
fit of the data in the region 
$\xi^+<\tilde{\xi}^+(\zeta^+>\tilde{\zeta}^+)$ by the
Boltzman distribution, \rule{0.3cm}{0.5mm}$\;$\rule{0.3cm}{0.5mm}$\;$\rule{0.3cm}{0.5mm}
 fit of the data
in the region $\xi^+>\tilde{\xi}^+(\zeta^+<\tilde{\zeta}^+)$ by the
formula $(1-\xi^+)^n$

\vspace{1cm}

\noindent Fig. 2: Angular distribution of $\pi^{\pm}$ - mesons in the 
reaction $\bar pp\to \pi^{\pm}X$ at 22.4~GeV/c (2a), 
\rule{1.3cm}{0.5mm} fit of the data in the region $\xi^+<\tilde{\xi}^+(\zeta^+>
\tilde{\zeta}^+)$ by the Bose-Einstein distribution, 
\rule{0.6cm}{0.5mm}$\,$.$\,$\rule{0.6cm}{0.5mm}
fit of the data in the region $\xi^+<\tilde{\xi}^+(\zeta^+>\tilde{\zeta}^+)$ by the
Boltzman distribution.\break
 $p^2_T$ - distribution of $\pi^{\pm}$ mesons in the
reaction $\bar pp\to \pi^{\pm}X$ at 22.4~GeV/c (2b), 
\rule{1.3cm}{0.5mm} fit of the data in the region $\xi^+<\tilde{\xi}^+(\zeta^+>
\tilde{\zeta}^+)$ by the Bose-Einstein distribution, 
\rule{0.6cm}{0.5mm}$\,$.$\,$\rule{0.6cm}{0.5mm}
fit of the data in the region $\xi^+<\tilde{\xi}^+(\zeta^+>\tilde{\zeta}^+)$ by the
Boltzman distribution, \rule{0.3cm}{0.5mm}$\;$\rule{0.3cm}{0.5mm}$\;$\rule{0.3cm}{0.5mm}
 fit of the data
in the region $\xi^+>\tilde{\xi}^+(\zeta^+<\tilde{\zeta}^+)$ by the
formula (20).

\vspace{1cm}

\noindent Fig. 3: Black body radiation intensity as a function of
dimensionless variable $x=\hbar\omega /kT$, 
\rule{0.3cm}{0.5mm}$\;$\rule{0.3cm}{0.5mm}$\;$\rule{0.3cm}{0.5mm}
$dE_{\omega}\sim x^2dx$ (Wien), 
\rule{0.6cm}{0.5mm}$\,$.$\,$\rule{0.6cm}{0.5mm} 
$dE_{\omega}\sim x^3e^{-x}$ (Rayleigh--Jeans), 
\rule{1.3cm}{0.5mm} $dE_{\omega}\sim x^3(e^x-1)^{-1}$ (Planck). 


\clearpage


\begin{thebibliography}{99}

\parskip=0pt

\bibitem{1} {A.A. Logunov, M.A. Mestvirishvili} and Nguen van {Hieu},
Phys. Lett. B25 (1967) 611
\par\noindent
\bibitem{2} {J. Benecke, T.T. Chou, C.N. Yang} and {E. Yen},
Phys. Rev. 188 (1969) 2159
\par\noindent
\bibitem{3} {R. Feynman}, Phys. Rev. Lett. 23 (1969) 415
\par\noindent
\bibitem{4} {V.A. Matveev, R.M. Muradyan} and {A.N. Tavkhelidze},
Lett. Nuovo Cim. 5 (1972) 907
\par\noindent
\bibitem{5} {A.M. Baldin}, Nucl. Phys. A447 (1985) 203
\par\noindent
\bibitem{6} {A.A. Baldin}, Phys. of Atomic Nuclei 56 (1993) 385
\par\noindent
\bibitem{7} {A.M. Baldin} and {L.A. Didenko}, Fortschritte der Phys.
38 (1994) 261
\par\noindent
\bibitem{8} {P.A.M. Dirac}, Rev. Mod. Phys. 21 (1949) 392
\par\noindent
\bibitem{9} {N. Ya. Vilenkin} and {Ya. A. Smorodinsky}, 
Sov. J. of Exp. and Theor. Physics, JETP 46 (1964) 1793
\par\noindent
\bibitem{10} {V.R. Garsevanishvili, V.G. Kadyshevsky, R.M. Mir-Kasimov}
and {N.B. Skachkov}, Sov. J. of Theor. and Math. Phys. 7 (1971) 203
\par\noindent
\bibitem{11} {H. Leutwyler}, Nucl. Phys. B76 (1974) 413
\par\noindent
\bibitem{12} {V.R. Garsevanishvili, A.N. Kvinikhidze, V.A. Matveev,
A.N. Tavkhelidze} and {R.N. Faustov}, Sov. J. of Theor. and Math. Phys.
23 (1975) 310
\bibitem{13} {V.R. Garsevanishvili and V.A. Matveev}, Sov. J. of Theor.
and Math. Phys. 24 (1975) 3
\par\noindent
\bibitem{14} {V.R. Garsevanishvili,} In: Proc. XIII International
Winter School of Theor. Phys., Karpacz, Poland, February, 1976, Ed.
J. Lukierski (Acta Universitatis Wratislaviensis, Wroclaw, 1976) p.315
\par\noindent
\bibitem{15} {D. Sivers, S.J. Brodsky and R. Blankenbecler}, Phys. Reports
23 (1976) 1
\par\noindent
\bibitem{16} {I. Schmidt and R. Blankenbecler}, Phys. Rev. D15 (1977) 3321
\par\noindent
\bibitem{17} {H. Leutwyler and J. Stern}, Ann. of Phys. (N.Y.) 112 (1978) 94
\par\noindent
\bibitem{18} {B.I.G. Bakker, L.A. Kondratyuk and M.V. Terentyev}, Nucl. Phys.
B158 (1979) 497
\par\noindent
\bibitem{19} {S.J. Brodsky}, In: Quarks and Nuclear Forces, Eds
{D.C. Fries} and {B. Zeitnitz}, Springer Verlag, Berlin, Heidelberg,
New York, 1982, p.81
\par\noindent
\bibitem{20} {V.A. Karmanov}, Sov. J. of Particles and Nuclei, 19 (1988) 525
\par\noindent
\bibitem{21} {D.S. Kulshreshtha and A.N. Mitra}, Phys. Rev. D37 (1988) 1268
\par\noindent
\bibitem{22} {V.R. Garsevanishvili} and {Z.R. Menteshashvili},
Relativistic Nuclear Physics in the Light Front Formalism, (Nova Science
Publishers, New York, 1993)
\par\noindent
\bibitem{23} {B. Despalques, V.A. Karmanov} and {J.-F. Mathiot},
Nucl. Phys. A589 (1995) 697
\par\noindent
\bibitem{24} {V.A. Karmanov} and {J.-F. Mathiot}, Nucl. Phys. A602
(1996) 388 \par\noindent
\bibitem{25} {J. Carbonell, B. Desplanques, V.A. Karmanov} and {
J.-F. Mathiot}, to appear in Phys. Reports.
\par\noindent
\bibitem{26} {B.S. Aladashvili} et al., Dubna-Ko\v sice-Moscow-Strasbourg-%
Tbilisi-Warsaw Collaboration.
Sov. J. of Nucl. Phys. 34(1981) 591
\par\noindent
\bibitem{27} {L.S. Azhgirey} et al., Phys. Lett. B387 (1966) 37
\par\noindent
\bibitem{28} K.G. Wilson, T.S. Walout, A. Harindranath, W.-M. Zhang, R.J. Perry
and St. D. Glazek, Phys. Rev. D49 (1994) 6720
\par\noindent
\bibitem{29} {V.G. Kadyshevsky, M. Mateev} and {R.M. Mir-Kasimov},
In: Proc. of the Intern. Seminar on Deep Inelastic and Inclusive Processes,
Sukhumi, 1975, Ed. by Inst. of Nucl. Research, Moscow.
\par\noindent
\bibitem{30} {A. Donkov, V.G. Kadyshevsky, M. Mateev} and {R.M.
Mir-Kasimov}, In: Proc. of {V.A. Steklov} Mathematical Institute, Moscow,
CXXXVI (1975) 85
\par\noindent
\bibitem{31} {E.G. Boos} et al., Nucl. Phys. B147 (1980) 45
\par\noindent
\bibitem{32} M. Klein, DESY Preprint, DESY 96-218, October, 1996, Invited
Talk at the 4-th-Intern. Conf. on Deep Inelastic Scattering, Rome, April, 1996.
\par\noindent
\bibitem{33} ALEPH Collaboration, CERN Preprint, CERN/PPE-96-999, to appear in Physics Reports.
\par\noindent
\bibitem{34} {R. Stock}, Plenary Talk at the XXVIII Intern. Conf. on High
Energy Physics, Warsaw, July, 25-31, 1996
\par\noindent
\bibitem{35} {K. Kadija}, Talk at the XXVIII Intern. Conf. on High
Energy Physics, Warsaw, July, 25-31, 1996
\par\noindent
\bibitem{36} {I. Derado}, Talk at the XIII Intern. Seminar on High
Energy Physics Problems, Dubna, September, 2-7, 1996
\par\noindent
\bibitem{37} {J. Bartke}, Talk at the XIII Intern. Seminar on High
Energy Physics Problems, Dubna, September, 2-7, 1996
\par\noindent
\bibitem{38} {B. Erazmus}, Talk at the XIII Intern. Seminar on High
Energy Physics Problems, Dubna, September, 2-7, 1996
\par\noindent
\bibitem{39} {E.L. Feinberg}, Sov. J. Uspekhi Fiz. Nauk, 164 (1971) 539
\par\noindent
\bibitem{40} {L.D. Landau} and {E.M. Lifshitz}, Statistical Physics
(Nauka, Moscow, 1976).
\par\noindent
\bibitem{41} {R. Hanbury-Brown} and {R.Q. Twiss}, Phil. Mag. 
45 (1954) 663~; Nature 178 (1956) 1046
\par\noindent
\bibitem{42} {G.I. Kopylov} and {Podgoretsky}, Sov. J. of Nucl. Phys.
19 (1974) 215
\par\noindent
\bibitem{43} {J. Pi\v sut} and {N. Pi\v sutova}, to appear in Physics
Letters B.
\par\noindent
\bibitem{44} {P. Danielewicz} and {G. Odyniec}, Phys. Lett. B157
(1985) 146
\par\noindent
\end{thebibliography}
\end{document}